\documentclass[aps,onecolumn,amssymb]{revtex4}
\usepackage{graphicx,color}
\usepackage{epsfig,amsmath}  

\def\p{\partial}
\def\bF{\bgroup \bf F\egroup}
\def\bx{\bgroup \bf x\egroup}
\def\bi{\bgroup \bf e\egroup}
\def\bn{\bgroup \bf n\egroup}
\def\vp{\varphi}
\def\tz{{\tilde z}}
\def\tphi{\tilde \phi}
\def\tq{{\tilde q}}

\def\sn{\mathop{\rm sn }\nolimits}
\def\const{\mathop{\rm const }\nolimits}
\def\be{\begin{equation}}
\def\ee{\end{equation}}
\def\red#1{\bgroup \color{red} #1\egroup}

\begin{document}

\title[Elastic instabilities in a layered cerebral cortex]{Elastic instabilities in a layered cerebral cortex: A revised axonal tension model for cortex folding}

\author{O. V. Manyuhina, David Mayett, and J. M. Schwarz}
\email{jschwarz@phy.syr.edu}
\affiliation{Department of Physics, Syracuse University, Syracuse, NY 13244, USA}

\begin{abstract}
We model the elasticity of the cerebral cortex as a layered material with bending energy along the layers and elastic energy between them in both planar and polar geometries. The cortex is also subjected to axons pulling from the underlying white matter. Above a critical threshold force, a ``flat'' cortex configuration becomes unstable and periodic unduluations emerge, i.e. a buckling instability occurs. These undulations may indeed initiate folds in the cortex. We identify analytically the critical force and the critical wavelength of the undulations. Both quantities are physiologically relevant values. Our model is a revised version of the axonal tension model for cortex folding, with our version taking into account the layered structure of the cortex. Moreover, our model draws a connection with another competing model for cortex folding, namely the differential growth-induced buckling model.  For the polar geometry, we study the relationship between brain size and the critical force and wavelength to understand why small mice brains exhibit no folds, while larger human brains do, for example.  Finally, an estimate of the bending rigidity constant for the cortex can be made based on the critical wavelength.  
\end{abstract}

\maketitle

\section{Introduction}
The cerebral cortex, or grey matter, is the outermost layer of nerve tissue covering the cerebrum and plays a key role in high-level cognitive functions, such as decision-making. The nerve cells in the cerebral cortex contain nonmyelinated axons, and the cortex is distinguished from the underlying nerve tissue consisting of nerve cells with myelinated axons, otherwise known as white matter. The geometry of the cortex varies across mammals~\cite{defelipe:2011}.  See Figure 1. In mice, the cortex is smooth, while in larger mammals, the cortex develops folds.  These folds allow for greater surface area of the cortex such that more neurons can participate in, and, therefore, presumably enhance higher-level cognitive functions. 

\begin{figure}[h]
\begin{center}
\includegraphics[height=60mm]{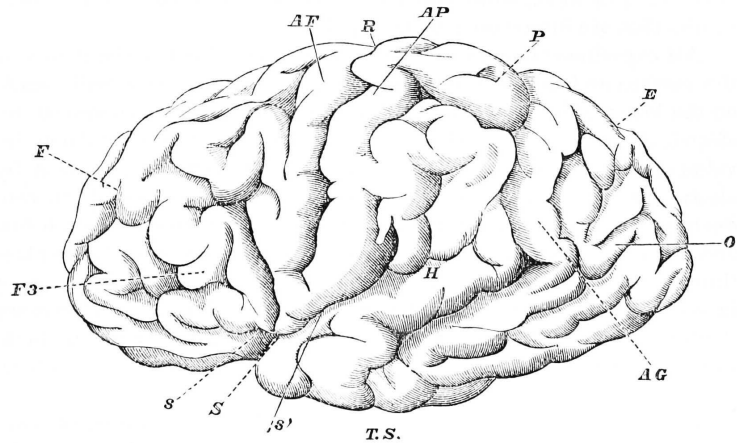}
\caption{Drawing of a human brain from an article by Sanger Brown M.D. in {\it Popular Science Monthly} {\bf 46}, 155 (1894). The typical radius of the human brain is $10$~cm.}
\end{center}
\end{figure}

To date, there are two competing mechanisms proposed to drive cortex folding. The first proposal claims the folds are driven by axonal tension from the underlying white matter drawing the sides of gyri (outward folds) together~\cite{vanessen}. See Figure 2a. This mechanism is appealing because it can be related to the efficient wiring of neurons via the minimization of distances. Moreover, this model does not invoke any elastic instabilities, i.e. buckling.  The second, competing proposal suggests that the folds are driven by buckling~\cite{richman}. See Figure 2b. More specifically, fast growth of the outermost layer of the cortex produces compressive stress that leads to buckling of this layer as modulated by the stiffness of the underlying foundation (comprised of the remaining layers of the cortex and the white matter).  Interestingly, this type of buckling model can also be invoked to study many shape changes in nature ranging from plant growth to geological folds~\cite{witten}.

\begin{figure}[th]
\centering
\includegraphics[width=0.9\linewidth]{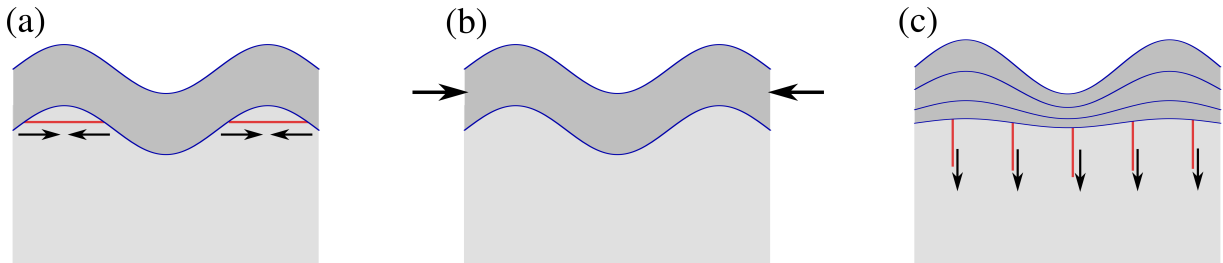}
\caption{Schematic of (a) the axonal tension model, which distinguishes between the cortex (denoted by the grey shading) and the underlying white matter, (b) the differential growth model, which distinguishes between the top layer of the cortex and the rest of the brain matter, (c) the model presented here.  The red lines denote axons, the black arrows denote the direction of the force.  Only three layers of the cortex are drawn for simplicity.}
\end{figure}

Indirect evidence for each mechanism exists. For instance, in fetal brains where most of the tissue below the cortex is surgically ablated prior to folds developing, folds eventually do develop~\cite{ablation}. This observation is typically invoked to demonstrate that the intracortical buckling drives folding and not axonal tension from the underlying white matter, though the effect of growth of cells outside the cortex, i.e. new white matter, cannot be ruled out~\cite{ablation}. In addition, a quantitative model of buckling of an elastic plate (the top layer of the cortex) supported by an elastic foundation (the white matter) yields a critical wavelength for buckling that agrees with the typical distance between folds, provided the Young's modulus of the white matter is 10 times less than the grey matter~\cite{richman}.  This assumption, however, has been called into question by recent indentation experiments of different parts of the cerebrum~\cite{dommelen}, showing that white matter is about 34\,\% stiffer than the grey matter.  Measurements of mechanical properties in vivo using magnetic resonance elastography at 100 Hz~\cite{manduca} also found that white matter is three times stiffer than grey matter.

For the axonal tension model, as originally formulated, neuronal pathways connecting gyri should be denser than those connecting sulci (inward folds). Some data supports this notion, though the results may be a matter of defining which surrounding regions belong to gyri and which belong to sulci~\cite{data1}. Moreover, cortical folds generated by linking different areas of the brain via axonal tension means that denser neuronal pathways should exhibit straighter white-matter trajectories.  There exists some correlation between denser neuronal pathways and straighter white-matter trajectories to support this notion~\cite{data2}. On the other hand, cuts in ferret brain tissue indicate that the tension does not run between gyri but radially outward~\cite{ferret}. Quantitative data for the axonal tension model at the same level as the buckling mechanism is currently lacking. 

Indeed, it could very well be that both mechanisms are at play in the folding of the cortex. If so, can we distinguish between the two?  To begin to do so, we develop a new model of the elasticity of the cortex that takes into account (1) the elongated, or rod-like, structure of nerve cells sitting in a ``background'' of softer, glial and progenitor cells and (2) the layered structure of the cortex~\cite{layer} (see Fig.~\ref{fig:layers}).  This combination of ingredients may make it reasonable to model the cortex as a layered liquid crystal with the neurons representing the liquid crystal molecules. 

With this ``cortex as a liquid crystal structure'' in what will turn out to be the smectic phase, we can revisit the axonal tension model and investigate the effect of pulling forces on the cortex. We will do this in both a planar geometry and a polar geometry and demonstrate that ``vertical pulling'' of the axons in the underlying white matter can lead to buckling in a layered structure. Our analysis allows for an updated version of the axonal tension hypothesis that is more consistent with the data. Prior to this work, all buckling models for cortex folding are based on ``horizontal compression''. 

The paper is organized as follows. The next section details our ``cortex'' as a smectic'' approach in a planar geometry to estimate the critical force needed to generate cortical folds with some critical wavelength.  Section III presents results for the polar geometry. Section IV summarizes our results and their implications.  

\section{Cortex as a smectic liquid crystal: Planar geometry}

The cortex consists of neurons, glial cells, and progenitor cells. The glial cells provide nutrients for the nerve cells they surround. Progenitor cells eventually become nerve cells (since nerve cells do not divide).  The shape of each nerve cell is rod-like with a cross-sectional diameter of order a micron and a length ranging from several hundred of microns to approximately a millimeter.  The mechanical rigidity along the axon of the nerve cell is provided for by microtubules. Microtubules are semiflexible polymers with a persistence length of approximately $1$~mm~\cite{microtubules}. Therefore, the nerve cells are rather rigid ``molecules''. The surrounding glial cells are softer~\cite{glial}. Since the glial cells are softer, we will assume that the elasticity of the cortex is dominated by the rigid, rod-like nerve cells.  

\begin{figure}[th]
\centering
\includegraphics[width=0.9\linewidth]{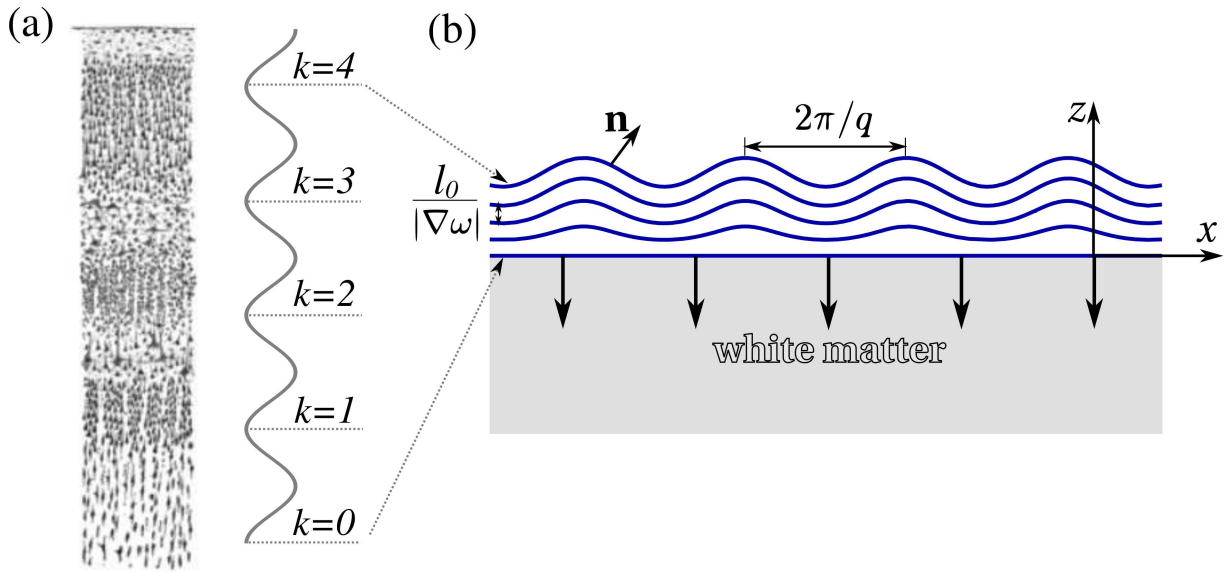}
\caption{\label{fig:layers}(a) A drawing of a Nissl-stained visual cortex of a human adult by Santiago Ramon y Cajal, showing a vertical cross-section, with the surface of the cortex at the top. Only the cell bodies (and not the elongated axons) are shown. (b)~Schematic of the planar model presented in Section 2 with notations used.}
\end{figure}

How are these rigid, rod-like nerve cells arranged in the cortex?  As indicated in Figure 3, they are predominantly oriented perpendicular to the outer surface of the cortex~\cite{defelipe:2011}. Moreover, the nerve cells in the cortex arrange themselves into six layers with the morphology differing slightly between layers.  For simplicity, we assume that all the layers are equivalent in thickness and in elastic properties. Given this extra spatial structure, we model the elasticity of the cortex as a smectic liquid crystal and then ask the following: What are the consequences of axons from the underlying white matter pulling vertically on the cortex in this planar geometry? The pulling of axons (nerve cells) has been well-established~\cite{pull} and given the orientation of axon highways in the underlying white matter~\cite{axonhighways}, vertical pulling is in keeping with observations.  So, as with the original version of the axonal tension model, here, the white matter enters the model solely via an applied strain due to active pulling stresses by the underlying axons and via boundary conditions. We will also include the effect of uniform cortical growth, as opposed to differential cortical growth, to begin to look for potential interplay between axonal pulling and growth driving cortex folding.

To quantify the effect of the applied vertical stain on the cortex, we consider the set of smectic surfaces $\omega(\bx,k)\equiv\bx\cdot\bn-kl=0$, or equivalently the peaks of density modulation $\delta\rho\propto\cos(2\pi\omega(\bx,k)/l)$ (see Fig.~\ref{fig:layers}). Here $\bx$ denotes the position in Cartesian coordinates $(x,y,z)$, $\bn$ is the unit normal to the layer, $k\in \mathbb Z$, and $l$ is the interlayer spacing~\cite{napoli:2009}. We will assume translational invariance in the $y$-coordinate such that the model is two-dimensional with the layers described by curves. The deformation of layers (curves), from the initial configuration described by $\bx_0=x\bi_x+z\bi_z$ with layer normal $\bn_0=\bi_z$ along $z$-axis, to the current configuration,  $\bx$, is characterised by the deformation gradient $\bF=\p\bx/\p\bx_0$. In this planar geometry, we assume the following mapping $x\mapsto (1+\alpha)x$ and $z\mapsto z+U(x,z)$, where $\alpha$ characterizes the lateral growth of the cortex, or the relative compressive strain such that $\alpha<0$, and  $U(x,z)$ is a displacement field due to vertical pulling of the axons. Then, the deformation gradient matrix and its inverse transposed in $\{\bi_x,\bi_z\}$ basis are
\be\label{eq:def}
 \bF=\begin{pmatrix}1+\alpha&0\\ \p_x U&1+\p_zU\end{pmatrix},\qquad \bF^{-T}=\frac 1{(1+\alpha)(1+\p_zU)}\begin{pmatrix}1+\p_z U&-\p_xU\\0&1+\alpha\end{pmatrix}.
\ee
Note that for generic elastic material it is hard to decouple the lateral and vertical displacements~\cite{landau} and, thus, define the deformation gradient ${\bf F}$. However, due to the layered structure along the $z$-direction, our cortex as a smectic behaves as a fluid within the layer ($x$-direction) and deforms as an elastic material in the $z$-direction. Therefore, the only physically feasible form of the coordinate transformation is stated above, which makes the matrix ${\bf F}$ triangular. Thus, the spatial gradient of isosurfaces in the current configuration can be computed as~\cite{napoli:2009} $\nabla \omega=\bF^{-T}\bn_0$, yielding
\be\label{eq:omeg}
\nabla\omega = -\frac{\p_x U}{(1+\alpha)(1+\p_zU)} \bi_x+\frac1{(1+\p_z U)}\bi_z.
\ee
The thickness of the deformed layers corresponds to $l_0/|\nabla\omega|$, where $l_0$ is the thickness of undeformed layers.  The elastic free energy density, accounting for the finite deformation such as compression and bending of layers, consists of two terms, respectively 
\be\label{eq:f}
f=\frac B2 (|\nabla \omega|^{-1}-1)^2+\frac K2 (\nabla\cdot\bn)^2,
\ee
where $B$ is compression modulus and $K$ is the bending rigidity. The ratio $\sqrt{K/B}$ defines the characteristic length scale of the order of the layer thickness $\simeq 1$~mm.

The displacement field, $U(x,z)$, can be further decomposed into a uniform dilation of layers along $z$-direction and an inhomogeneous displacement, $u(x,z)$, so that $U(x,z)=\gamma z+u(x,z)$.  Assuming $u\ll 1$ and expanding the layer dilation and the unit normal we find
\begin{eqnarray}\label{eq:interl}
\frac 1{|\nabla \omega|}-1\simeq \gamma +\p_z u - \frac{1+\gamma}2\frac{(\p_xu)^2}{(1+\alpha)^2},\\
\bn=\frac{\nabla\omega}{|\nabla \omega|}\simeq-\frac{\p_x u}{(1+\alpha)}\bi_x+\bi_z.
\end{eqnarray}
To recover an equilibrium spacing and to compensate for the applied strain $\gamma$, the layers should bend locally $\p_x u\neq0$, so that Eq. \ref{eq:interl} tends to zero~\cite{singer:1993}. The regularity of the wavelength of brain folds allows us to assume the selection of the certain wavelength and thus look for the periodic solution, $u(x,z)=\phi(z)\cos(qx)$. Replacing this {\it ansatz} into Eq.~\ref{eq:f}  and integrating over the period $2\pi/q$ we arrive at the free energy, 
\be\label{eq:ene}
{\cal F}= \frac{hB}4\int_0^1 d\tz\,\bigg[2\gamma^2+(\p_{\tz}\tphi)^2 + \tphi^2\tq^2\underbrace{[\kappa^2\tq^2- \gamma(1+\gamma)]}_{\gtrless} +\frac{3 (1+\gamma)^2\tq^4 }{16}\tphi^4  \bigg],
\ee
where we introduced the dimensionless variables 
\be\label{eq:dimen}
\kappa^2=\frac {K(1+\alpha)^2}{B h^2},\quad \tz= \frac z h,\quad \tq=\frac {q h}{1+\alpha},\quad \tphi=\frac \phi h.
\ee
This free energy has not previously been studied beyond the harmonic limit~\cite{napoli:2009}. 

As for parameters, the thickness of the cortex scales does not vary much among the mammals, namely $h\simeq 2-5$~mm~\cite{layer}; the elastic modulus varies in the range $B\simeq0.1-2$~kPa~\cite{dommelen,elastic}, while we are not aware of any measurement of the bending rigidity $K$.  Note that the uniform growth factor $\alpha$ is absorbed in the coefficients $\kappa$ and dimensionless wavenumber $\tq$.

In equilibrium, we require the vanishing of the first variation of the free energy~\ref{eq:ene} $\delta{\cal F}=0$, yielding the Euler--Lagrange equation  and the boundary conditions. The upper interface with surrounding fluid ($\tilde z=1$)  is free, thus $\p_{\tz}\tphi|_{\tz=1}=0$. The boundary condition at the lower grey--white  matter interface ($\tilde z=0$) depends mainly on the difference of mechanical properties between white and grey matter. If  the white matter is softer, then the interface ($\tilde z=0$) is free and we may ignore the variation of the displacement field along the thickness of the cortex $h$ ($(\p_{\tz}\tphi)^2\ll \tphi^2\tq^2$). Then, any infinitesimal strain $\gamma$ leads to an instability of the flat layers towards a periodically modulated state ($\tphi\neq0$) with the wavenumber $\tilde q\lesssim\sqrt{\gamma(1+\gamma)}/\kappa$, which follows from  Eq.~\ref{eq:ene}.  This picture is rather na\"ive, and contradicts recent experimental measurements, where white matter was found to be significantly stiffer than the grey matter~\cite{dommelen, manduca}. Therefore, we assume that the grey--white matter interface ($\tilde z=0$) remains flat $\tphi|_{\tz=0}=0$, though the physiological boundary condition is not yet known.

Similar to the Helfrich--Hurault instability in nonliving smectic liquid crystals~\cite{napoli:2009,singer:1993,helfrich,hurault} we expect, that above some critical threshold $\gamma_{\rm cr}$, undulations of layers are energetically favoured to minimize the compression energy in expense of the bending. By bending locally with a slope $\p_x u\sim \phi q\neq 0$ the layers tend to recover the equilibrium spacing and decrease the strain $\gamma$~\ref{eq:interl}. More precisely it follows from Eq.~\ref{eq:ene} that if  $\gamma(1+\gamma)>\kappa^2\tq^2$ there exists a non-trivial solution ($\tphi\neq0$ and $\tq\neq0$) extremizing the free energy~\cite{singer:1993}. This is necessary but not sufficient condition for the buckling profile to be favoured. Below we derive the stability criterion, which relates the control parameter $\gamma$ and the wavelength $L_x=2\pi/q$ with thickness $h$ and elastic moduli $\sqrt{K/B}$, which are intrinsic parameters of the system.

\begin{figure}[th]
\centering
\includegraphics[width=0.9\linewidth]{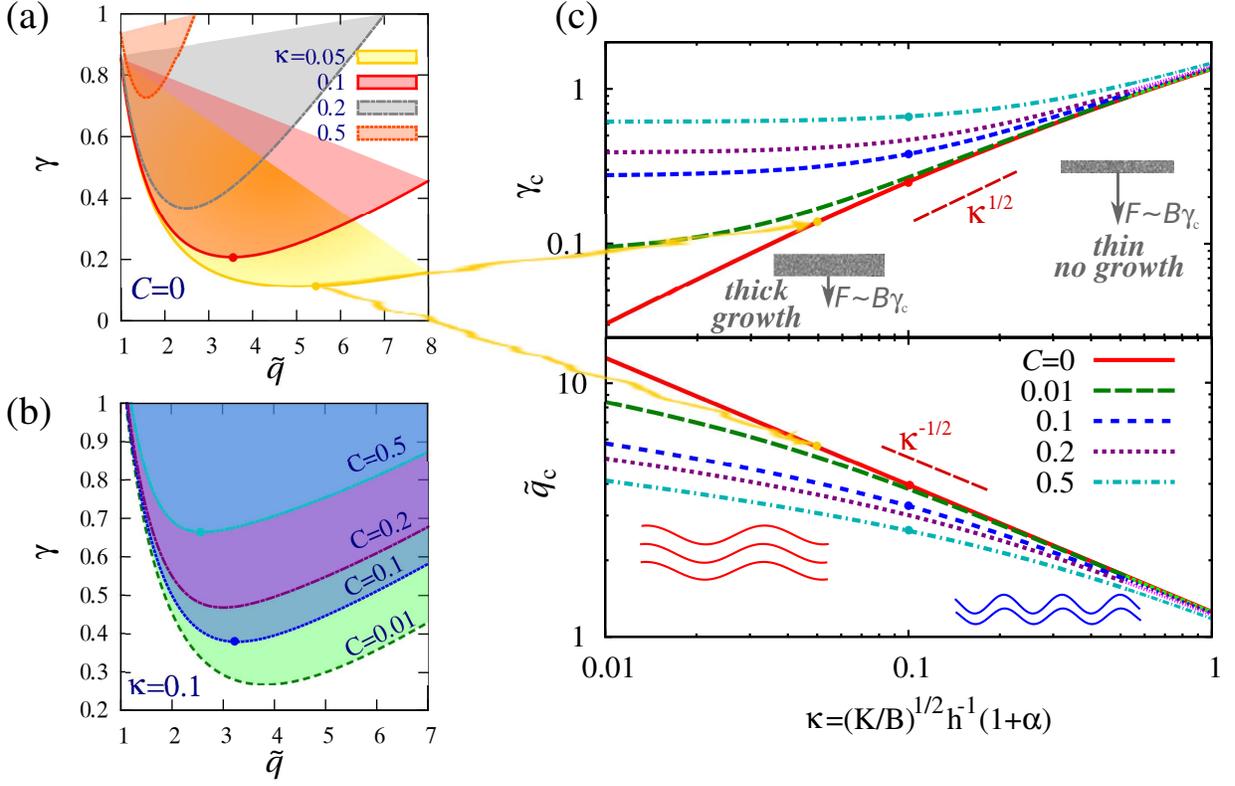}
\caption{\label{fig:nu}(a) The  condition $\sqrt{\mu}=\pi/2$ (${\cal C}=0$) for different values of $\kappa$ plotted in $\gamma$--$\tq$ plane with the minimum of the curves corresponding  to the threshold, Eq.~\ref{eq:thresh}; (b) The condition Eq.~\ref{eq:nu} for different values of ${\cal C}$ and $\kappa=0.1$; (c) Plot of the critical points $\{\gamma_c,\tq_c\}$ as a function of $\kappa$ for different values of ${\cal C}$. The critical strain $\gamma_c$ decreases for a thicker cortex and in presence of growth ($\alpha<0$, small $\kappa$). The dependence of the critical wavelength $L_x=2\pi\sqrt{K/B}/(\kappa{\tilde q}_c)$ on $h$ is shown in Figure 6.}
\end{figure}

The first integral of the Euler--Lagrange equation associated with Eq.~\ref{eq:ene} is
\be\label{eq:phi}
(\p_\tz\tphi)^2={\cal V}(\tilde\phi), \quad {\cal V}(\tilde\phi)\equiv\frac 3{16}(1+\gamma)^2\tq^4\tilde \phi^4+\tilde \phi^2\big[-\gamma(1+\gamma)\tq^2+\kappa^2\tq^4\big]+{\cal C}.
\ee
The integration constant ${\cal C}$ can be related to the active stresses exerted by axons $\sigma_{\rm axon}$, assuming continuity of the normal stress at the white--grey matter interface ($\tz=0$). The latter can be found from the variation in the free energy~\ref{eq:f}, $\delta f\propto-\sigma_{ij}\p_j u_i$~\cite{landau}, yielding $\sigma_{\rm axon}=\sigma_{zz}\propto B\sqrt{\gamma^2 +C/2}$. We may also think of ${\cal C}$ as an amplitude of perturbation of $\tphi$, since in~\ref{eq:phi} we have $\tphi|_{\tz=0}=0$. As ${\cal C}\to 0$ we are at the instability threshold, while above the threshold the amplitude $\sqrt{\cal C}$ is finite but small because of the truncated series expansion of the free energy~\ref{eq:ene}, hence we investigate ${\cal C}<1$. Rewriting the potential on the RHS of Eq.~\ref{eq:phi} as ${\cal V}(\tilde \phi)=\frac{3}{16}(1+\gamma)^2\tq^4(\tphi^2-\phi_1)(\tphi^2-\phi_2)$ and intergrating, we find the general solution in terms of the Jacobi elliptic function, or 
\be\label{eq:sn}
\tilde \phi(\tilde z)=\sqrt{\phi_1}\sn\bigg(\nu\,\tz,\,\frac{\phi_1}{\phi_2}\bigg),\qquad \nu=\frac{\sqrt 3}4(1+\gamma)\tq^2 \sqrt{\phi_2},
\ee
which satisfies the boundary condition $\tilde \phi|_{\tilde z=0}=0$. The period of Eq.~\ref{eq:sn} is $4{\cal K}(\phi_1/\phi_2)/\nu$, where ${\cal K}$ is the complete elliptic integral of the first kind. The condition at the upper interface, corresponding to the maximum of the displacement field, determines the threshold criterion which is 
\be\label{eq:nu}
\nu={\cal K} \bigg(\frac{\phi_1}{\phi_2}\bigg),\quad \phi_{1,2}=\frac8 3\frac{\mu\mp\sqrt{\mu^2-3{\cal C}(1+\gamma)^2\tq^4/4}}{(1+\gamma)^2\tq^4},
\ee 
with $\mu=\tq^2(\gamma+\gamma^2-\kappa^2\tq^2)$. In the limiting case ${\cal C}\to 0$, shown in Fig.~\ref{fig:nu}a, the above condition falls into $\sqrt{\mu}=\pi/2$ with the threshold being written explicitly as
\be\label{eq:thresh}
\gamma_c^0=\frac 12 \big(-1+\sqrt{1+4\pi\kappa}\big), \qquad \tq_c^0=\sqrt{\frac \pi 2}\sqrt{\frac 1\kappa}. 
\ee
This result coincides with the harmonic case studied in~\cite{napoli:2009}, modulo a factor of two in $\gamma^0_c$.  Note that in this case, the wavelength of the instability, $L_x^0=2\pi/q_c=2\sqrt{2\pi K/B}$ (Eq.~\ref{eq:dimen}), is given by the ratio of elastic constants and neither depends on the thickness $h$ nor on the lateral growth. On the contrary, the threshold $\gamma_c^0$ (Eq.~\ref{eq:thresh}), the minimum of the curve in Fig.~\ref{fig:nu}a and Fig.~\ref{fig:nu}c for ${\cal C}=0$, is lower in the absence of growth $\alpha\neq 0$ and for higher thickness $h$ of the cortex (small $\kappa$). This means that for thicker cortices the instability is more likely to happen, assuming the same elastic constants $K/B$ and growth rate. 

Now that we know $\gamma^0_c$, is this a physiologically accessible value? To answer this question, we can estimate the vertical stress exerted by axons at the white--grey matter interface, $\tz=0$, required for the instability to happen since $\sigma_{\rm axon}\simeq B\gamma_c^0$. The typical value of elastic modulus of cortical neurons is approximately $200$~Pa~\cite{elastic}, which is related to $B$. The value of bending rigidity $K$ is not available in the literature, however, based on the analogy with smectic liquid crystals we assume $\sqrt{K/B}\sim 0.1-1$~mm is of the order of the layer thickness.  Thus, for a human brain with $h\simeq 4$~mm we can estimate $\kappa\sim0.025-0.25$ (assuming no growth), yielding the threshold $\gamma_c^0\simeq 0.07-0.5$ with necessary stress $\sigma_{\rm axon}\sim 10-100$~Pa, or the force of $10-100$~pN per the unit area of $1~\mu$m$^2$. This prediction is consistent with tension measurements of neurons~\cite{pull}.

In general case, with  $(\p_\tz\tphi)^2|_{\tz=0}={\cal C}\neq0$, we are above the threshold in Eq.~\ref{eq:thresh}, and the axons exert higher stresses than $\sigma_{\rm axon}\gtrsim B\gamma_c^0$. 
In Fig.~\ref{fig:nu}b, we plot threshold curves (Eq.~\ref{eq:nu}) in the $\gamma$--$\tq$ plane for $\kappa=0.1$ and different values of integration constant ${\cal C}$.  We  identify the threshold (the minimum) numerically and show  the values $\{\gamma_c,\tq_c\}$ in Fig.~\ref{fig:nu}c as function of the dimensionless variable, $\kappa$. As expected, $\gamma_c>\gamma_c^0$, and the wavelength $L_x\sim 1/\tq_c$ increases above the threshold. Note that Ref.~\cite{singer:1993} obtains the opposite trend.  This difference is due to the additional $(1+\gamma)$ factors in our free energy, where we have not assumed that $\gamma$ is small. 

Before concluding this section, let us address this instability for different mammalian species. Assuming that elastic constants are of the same order for different species~\cite{species}, we expect that for smaller $h$, $\kappa$ increases, thus, the threshold, $\gamma_c$, also increases (see Fig.~\ref{fig:nu}c). Plausibly, in small species (small $h$) not enough force is generated by axons to overcome the threshold such that no folds emerge. On the other hand, the cortex thickness scales logarithmically with brain size~\cite{brainsize} so that we should analyse the role of geometry and confinement on the instability threshold for layer buckling before drawing conclusions across species.

\section{Cortex as a smectic liquid crystal: Polar geometry}

Now, we model the cortex as 2D set of curves in polar coordinates $(\vp,r)$, confined between two radii $r\in[R_0;R_0+h]$ and $\vp\in[0;2\pi]$. Here, $R_0$ is the lateral size of an idealised circular brain, which varies among mammalian species from $5$~mm to $20$~cm~\cite{brainsize}. The layers in the ground state are concentric circles radial with the position $\bx_0=r \bi_r$, and the normal to the layers $\bn=\bi_r$. Neglecting the effect of growth in this section, we consider a deformation map $\vp\mapsto \vp$ and $r\mapsto r+v(r,\vp)$ with the corresponding deformation gradient
\be\label{eq:def1}
\bF=\begin{pmatrix}1&\quad 0 \\ \p_\vp v/r & \quad 1+\p_r v\end{pmatrix},\quad\bF^{-T}=\frac{1}{(1+\p_r v)}\begin{pmatrix}1+\p_r v & -\p_\vp v/r\\ 0 & \quad 1\end{pmatrix},
\ee
Note that we do not incorporate growth here since it does not drive the instability found in the previous section. Then, the spatial gradient of isosurfaces in the current/target configuration can be computed as $\nabla \omega=\bF^{-T}\bi_r$, or 
\be
\nabla\omega=-\frac{\p_\vp v}{r(1+\p_r v)} \bi_\vp+\frac1{(1+\p_r v)}\bi_r.
\ee
The thickness of the deformed layers corresponds to $l_0/|\nabla\omega|$, where $l_0$ is the thickness of undeformed layers.  The normal to the layers and its divergence, i.e. curvature, are given by
\be
\bn=\frac{\nabla\omega}{|\nabla \omega|}, \qquad \nabla\cdot \bn=\frac{\p_r(r n_r)}r+\frac {\p_\vp n_\vp}r\simeq\frac  1 r -\frac {\p_{\vp\vp} v}{r^2},
\ee 
where we have linearized $n_r\simeq 1$ and $n_\varphi\simeq -\p_\varphi v/r$ for small deformations $v\ll 1$. The free energy density in polar coordinates can now be obtained by inserting the above expressions into Eq.~(\ref{eq:f})
\be\label{eq:polar}
{\cal F}_{\rm pol}=\frac 12 \int_0^{2\pi}d\varphi\int_{R_0}^{R_0+h} \!\!\!dr\,r B\bigg(\p_r v -\frac{1+\p_r v}2\frac{(\p_\varphi v)^2}{r^2}\bigg)^2+K \bigg(\frac 1 r-\frac{\p_{\varphi\varphi} v }{r^2}\bigg)^2.
\ee
Similar to the previous section we are looking for solutions in the form $v(r,\vp)=\gamma r +\psi(r)\cos (q_{\varphi}\vp)$. First we consider free interfaces and assume no radial dependence of the perturbation ($\psi(r)=\const$). After  integration we find the following condition for  the wavenumber $q^2_\varphi\lesssim \gamma(1+\gamma)\log(1+\eta)(1+\eta)^2/(\xi^2\eta(2+\eta))$, which depends on both the elastic constants $\xi^2=K/(BR_0^2)$, and dimensionless thickness $\eta=h/R_0$.

\begin{figure}[th]
\centering
\includegraphics[width=0.9\linewidth]{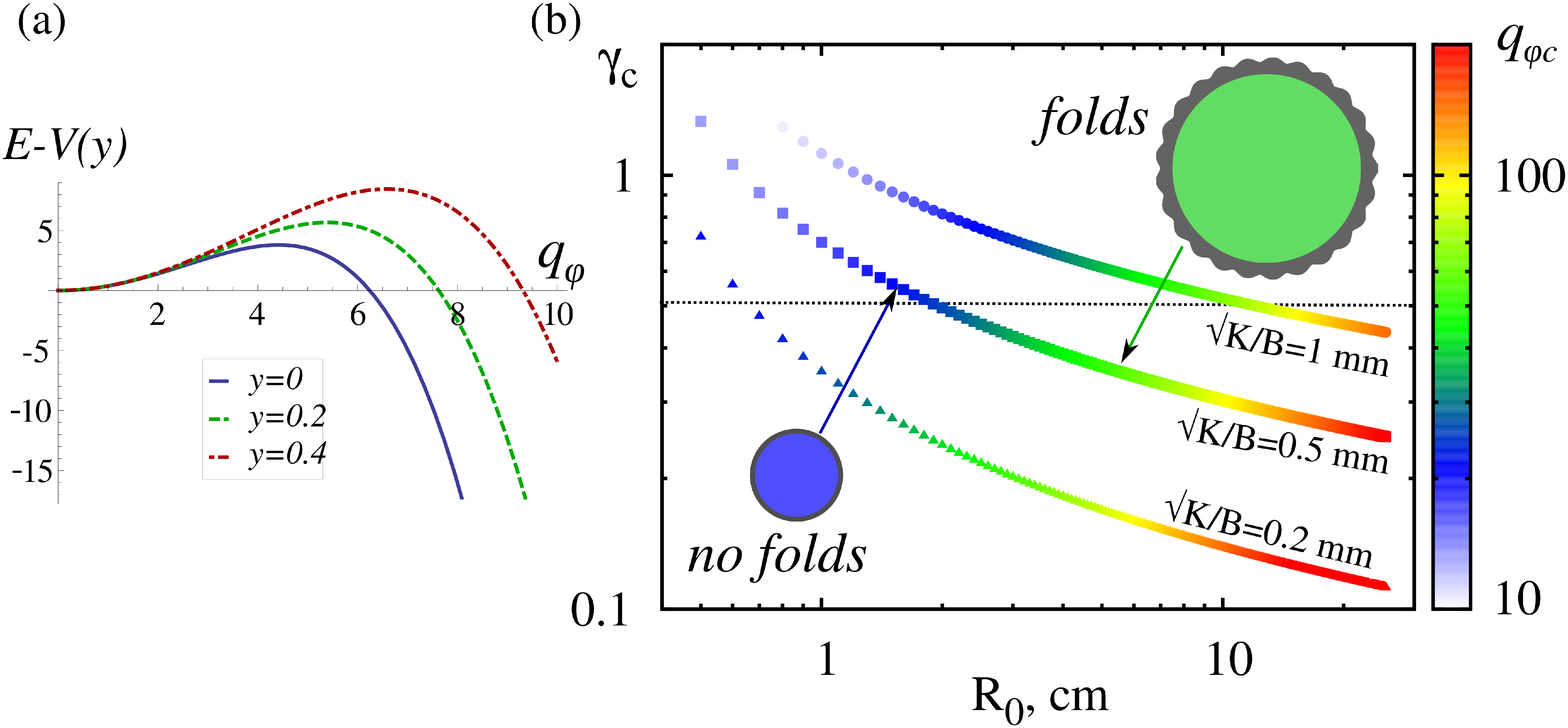}
\caption{\label{fig:wkb}(a) $E-V(y)$ as function of the wavenumber $q_\varphi$ for different distances $y$ with $\xi=0.1$ and $\gamma=0.3$. (b) The critical parameters associated with Eq.~\ref{eq:curv} as a function of the brain size $R_0$, assuming the thickness $h$ of the cortex depends logarithmically on $R_0$. For the human brain, $R_0\simeq 10$~cm and $h\simeq 4$~mm. The dotted line denotes the critical strain (or force) exceeding reported values of axonal tension such that the initiaton of folds due to axonal tension would not be observed.}
\end{figure}

Accounting for the difference in boundary conditions (the inner interface $r=R_0$ is clamped, while the outer interface $r=R_0+h$ is free) so that $\p_r\psi\neq0$, and integrating out the $\vp$-dependence in Eq.~\ref{eq:polar} we get
\be
{\cal F}_{\rm pol}[\tilde\psi]\propto \int_0^{\log (1+\eta)}\!\!\!\!dy\, \Big\{(\p_y\tilde\psi)^2+\tilde \psi^2\big[\xi^2 q_\varphi^4 e^{-2y}-\gamma(1+\gamma)q_\varphi^2\big] \Big\} +O(\tilde\psi^4),
\ee
where $y=\log(r/R_0)$, and $\tilde \psi=\psi/R_0$. The equilibrium equation for $\tilde\psi$ is a Schr\"odinger-type equation, describing a particle with energy $E$ in the potential well $V(y)$ (see Fig.~\ref{fig:wkb}a), written as
\be
\tilde\psi''= (V(y) -E)\tilde \psi,\qquad V(y)=\xi^2q_\varphi^4e^{-2y} ,\quad E=\gamma(1+\gamma)q_\varphi^2.
\ee

Assuming the WKB approximation for the classical region $E>V(y)$~\cite{bender:book}, we obtain 
\be
\tilde \psi_{\rm WKB}(y) =C_{\pm} (E-V(y))^{-1/4}\exp\Big\{\pm i \int_y dt\sqrt{E-V(t)}\Big\}.
\ee
Satisfying the boundary conditions $\tilde \psi_{\rm WKB}|_{y=0}=0$ and $\p_y\tilde \psi_{\rm WKB}|_{y=\log(1+\eta)}=0$, we find a relationship similar to Eq.~\ref{eq:nu}, which reads as
\be\label{eq:curv}
\int_0^{\log{(1+\eta)}}\kern-10pt dy\,\sqrt{E-V(y)}+ \arctan\bigg(\frac{q_\varphi \xi^2(1+\eta)}{2 [\gamma(1 +\gamma)(1+\eta)^2-q_\varphi^2 \xi^2]^{3/2}}\bigg)=\frac \pi 2.
\ee
In fact, the first term can be integrated and cast in the closed form using 
\begin{eqnarray}
\int \sqrt{E-V(y)}=q_\varphi \sqrt{\gamma  (1+\gamma )} \big[\log\big(\gamma +\gamma ^2+\sqrt{\gamma  (1+\gamma ) (\gamma +\gamma ^2-q_\varphi^2 \xi ^2)}\big) \nonumber\\
-\sqrt{1-q_\varphi^2 \xi ^2/(\gamma +\gamma ^2)}\big].
\end{eqnarray} 
More importantly, the curves defined by Eq.~\ref{eq:curv} have minima in the $\gamma$--$q_\varphi$ plane. These minima determine the threshold $\gamma_c$.

For the following analysis, we assume a logarithmic dependence of $h\in[0.5:5]$~mm on the size of the brain $R_0\in[0.5:25]$~cm~\cite{brainsize}.  We observe that $\gamma_c$ decreases with the increasing system size $R_0$, while the number of undulations $q_{\varphi c}$ increases as shown in Fig.~\ref{fig:wkb}b. In other words, the smaller the brain, the less likely it is for cortex folds to develop given the increased threshold. Moreover, even if the threshold were met, the distance between folds would be larger such that there would be fewer folds. Our results may explain why mice brains do not exhibit folds, while human brains do. To be more precise, for a typical human brain with $R_0\sim10$~cm, we obtain the typical dimensionful wavelength $L_x=2\pi R_0/q_{\varphi c}$ of the order of $1$~cm for $\sqrt{K/B}\simeq 1$~mm ($\xi=0.01$) and $4$~mm for $\sqrt{K/B}=0.2$~mm ($\xi=0.002$, or 25 times smaller $K$) as shown in  Fig.~\ref{fig:wkb}b with the colored sidebar. For a typical mouse brain, $R_0\sim 1~$cm such that the typical wavelength is about $6$~mm (for $\xi=0.1$). However, the critical strain required to initiate the instability exceeds unity such that the instability would not be accessible.  We should also note that the value of $\gamma_c$ related to the stress exerted by the axons, $\sigma_{rr}\sim B\gamma_c$, and depends strongly on the thickness $h$ of the cortex.

\section{Discussion}

Given the two dominant pre-existing mechanisms of cortical folding (axonal tension and buckling), our cortex-as-a-smectic approach represents a novel way to think about the elasticity of the cortex. With this approach we demonstrate that vertical pulling forces via axonal tension can lead to buckling in the cerebral cortex.  Prior buckling models of the initiation of cortex folding are a consequence of horizontal compression due to growth of the outermost layer of the cortex~\cite{richman,bayly1}, with the original version assuming that the white matter is much softer than the cortex, contrary to observation~\cite{dommelen,manduca}.  Our revised version of the axonal tension idea is in keeping with the observation that the white matter is stiffer than the cortex~\cite{dommelen, manduca} and that neurons in the white matter just beneath the cortex are oriented perpendicularly to the cortex~\cite{axonhighways}.  While some doubt has been cast on the original version of the axonal tension model since circumferential tension along the axes of gyri (from one side of the ``hill'' to other side) is not observed, but radial tension is~\cite{ferret}. Our model does not conflict with this observation. The observation of circumferential tension near the bases of sulci (the valleys) presumably sets in a later stage in the folding process~\cite{ferret}.  Here, we have focused on the initiation of the folding process in planar and polar geometries. 

With our simple model, we analytically predict the critical wavelength and critical strain of axon pulling as a function of cortex thickness and elastic properties. In the planar geometry, we find that the critical strain increases with uniform growth, while the critical wavelength does not depend on growth but only on the elastic properties and cortex thickness, $h$. By investigating the polar geometry, we address cortex folding across mammals. We find that smaller brains (smaller $h$) require a larger strain/stress to initiate buckling/folding and that the critical wavelength increases with brain size.  While we did not investigate the effect of growth in the polar geometry since the instability presented here is not driven by growth, it would be interesting to extend this case to include growth, particularly, differential growth, or stress-dependent growth~\cite{bayly1}.  

\begin{figure}[th]
\centering
\includegraphics[height=7cm,width=11cm,clip=true]{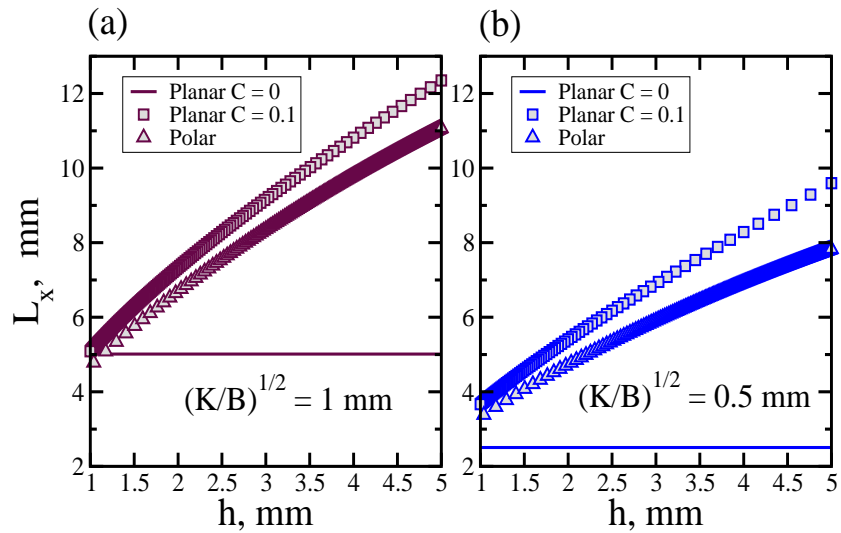}
\caption{\label{fig:thickness}(a) Plot of the critical wavelength $L_x$ as a function of thickness $h$ for the planar geometry with ${\cal C}=0$, the planar geometry with ${\cal C}>0$, and the polar geometry for comparison. Here, $\sqrt{K/B}=1$~mm. (b) Same as in (a) except with $\sqrt{K/B}=0.5$~mm.}
\end{figure}

In Figure~\ref{fig:thickness} the critical wavelength $L_x$, is plotted as a function of cortex thickness $h$ for two different values of $\sqrt{K/B}$ for both geometries considered here. Given the geometry of the human brain, Fig.~\ref{fig:thickness}a is more in keeping with observations than Fig.~\ref{fig:thickness}b.  Our results therefore indicate that $\sqrt{K/B}\sim 1$~mm is consistent with observations~\cite{defelipe:2011}, suggesting an urgency for direct measurements of the bending rigidity of the cortex $K$ to test this prediction. 

Most of the cortical folds are simple folds---a simple ``indentation'', if you will, though some folds exhibit more structure. For instance, there exist secondary folds deeper inside the brain.  We will call these more complicated structures, T-folds. And while such folds are more rare, a complete theory of cortex folding should be able to explain such emergent structures. To obtain these structures, growth, more details of the underlying white matter, and possibly constraints~\cite{creases} will have to be incorporated into the model. 

Finally, we have focused on the material properties of the cerebral cortex to better understand its shape. However, how does such properties affect its function?  In developing a more accurate theory of the ``brain as a material'', can we better understand its function?  For example, it would be interesting to couple elasticity with connectivity models of the cortex~\cite{houzel} to determine more precisely the interplay between structure and function in the brain.

Acknowledgements: OVM acknowledges stimulating discussion with Gaetano Napoli. The authors acknowledge comments from a referee on the justification of boundary conditions. The authors acknowledge financial support from the Soft Matter Program at Syracuse University. 

\section*{References}

\bibliographystyle{unsrt}
\bibliography{ref_cortex}

\end{document}